\documentclass[
 reprint,
superscriptaddress,
 amsmath,amssymb,
 aps,
 prl,   
 longbibliography,
floatfix, 
]{revtex4-2}

\usepackage{graphicx}
\usepackage{dcolumn}
\usepackage{bm}

\usepackage{float} 
\usepackage{xcolor}
\usepackage{amsmath}
\usepackage{mathrsfs} 

\newif\ifdraft
\drafttrue  

\begin{document}


\title{2D Canonical Approach for Beating the Boltzmann Tyranny Using Memory}

\author{Rafael Schio Wengenroth Silva}
\affiliation{Departamento de Física, Universidade Federal de São Carlos, 13565-905 São Carlos, SP, Brazil}

\author{Soumen Pradhan}
\affiliation{Julius-Maximilians-Universität Würzburg, Physikalisches Institut and Würzburg-Dresden Cluster of Excellence ct.qmat, Lehrstuhl für Technische Physik, Am Hubland, 97074 Würzburg, Deutschland}

\author{Fabian Hartmann}
\affiliation{Julius-Maximilians-Universität Würzburg, Physikalisches Institut and Würzburg-Dresden Cluster of Excellence ct.qmat, Lehrstuhl für Technische Physik, Am Hubland, 97074 Würzburg, Deutschland}

\author{Leonardo K. Castelano}
\affiliation{Department of Physics, Federal University of São Carlos, 13565-905 São Carlos, SP, Brazil}

\author{Ovidiu Lipan}
\affiliation{Department of Physics, University of Richmond, 28 Westhampton Way, Richmond, Virginia 23173, USA}

\author{Sven Höfling}
\affiliation{Julius-Maximilians-Universität Würzburg, Physikalisches Institut and Würzburg-Dresden Cluster of Excellence ct.qmat, Lehrstuhl für Technische Physik, Am Hubland, 97074 Würzburg, Deutschland}

\author{Victor Lopez-Richard}
\affiliation{Departamento de Física, Universidade Federal de São Carlos, 13565-905 São Carlos, SP, Brazil}
\email{vlopez@df.ufscar.br}

\date{\today}

\begin{abstract}
The 60 mV/decade subthreshold limit at room temperature, coined as the Boltzmann tyranny, remains a fundamental obstacle to the continued down-scaling of conventional transistors. While several strategies have sought to overcome this constraint through non-thermal carrier injection, most rely on ferroelectric-based or otherwise material-specific mechanisms that require complex fabrication and stability control. Here, we develop a universal theoretical framework showing that intrinsic memory effects in nanometric field-effect transistors can naturally bypass this limit. Within the Landauer–Büttiker quantum transport formalism, we incorporate charge-trapping mechanisms that dynamically renormalize the conduction band edge. The resulting analytical expression for the subthreshold swing explicitly links memory dynamics to gate efficiency, revealing that a reduced carrier generation rate or enhanced trapping activity leads to sub-thermal switching, thus breaking the Boltzmann barrier. The model captures key experimental features and provides clear, generalizable design principles, establishing memory-assisted transistors as a robust pathway toward ultra–low-power and multifunctional electronic architectures.
\end{abstract}

\maketitle


The relentless pursuit of energy-efficient computing has reached a critical juncture as conventional silicon technology approaches fundamental thermodynamic limits. The so-called ``Boltzmann tyranny'', which imposes a minimum subthreshold swing (SS) of $\sim$60 mV/decade at room temperature, arises from the Boltzmann tail of Fermi-Dirac statistics governing carrier injection via thermionic emission~\cite{Wang2019}. Overcoming this constraint is essential for scaling devices to smaller dimensions, reducing supply voltages, and extending Moore’s law~\cite{Lukyanchuk2022}. A steeper swing also enables far more energy-efficient computation in integrated circuits~\cite{Qiu2018}. As computational demands continue to increase, breaking this thermodynamic barrier has become a key challenge for sustaining the progress of semiconductor technology~\cite{Leiserson2020}.

To address this challenge, various steep-slope device concepts have been investigated~\cite{Zhai2021}. These include dielectric engineering~\cite{Ilatikhameneh2015}, tunneling FETs employing metallic strips~\cite{Kanagarajan2024} but limited by leakage~\cite{Xiao2024}, cold-source FETs that tailor the source density of states~\cite{Gan2020,Lyu2020}, and Dirac-source devices exploiting two-dimensional materials~\cite{Qiu2018,Xiao2020,Tang2021}. Among them, negative-capacitance FETs stand out for incorporating ferroelectric layers to achieve potential amplification~\cite{Salahuddin2008}, particularly when combined with two-dimensional semiconductors that offer improved electrostatics and multifunctional integration~\cite{Salahuddin2008,Yang2025,Chattopadhyay2024,Kamaei2023,Wang2022}. Indeed, hysteresis-free sub-60 mV/decade switching has been demonstrated in MoS$_2$ channels~\cite{Si2018,Cho2021,Tu2020}, and theoretical proposals further highlight alternative routes such as Rashba-driven topological phase transitions~\cite{Nadeem2021}. Experimentally, steep swings as low as 2.3 mV/decade have been reported in WSe$_2$ via impact ionization~\cite{Choi2022}, while vertical nanowire heterojunctions have demonstrated operation below 30 mV/decade~\cite{Shao2025}.

Beyond these approaches, memory-enhanced transistors (memtransistors) could emerge as a powerful alternative for achieving steep-slope behavior by exploiting nonequilibrium charge dynamics rather than relying on ferroelectric-based gate stacks~\cite{NatureNanotechnologyEditorial2020}. Unlike conventional transistors governed solely by electrostatics, or ferroelectric devices limited by material hysteresis, our configuration of memtransistors utilize dynamic internal states such as charge trapping and release to modulate the channel potential in real time~\cite{Yang2024,Aziz2022}. This mechanism enables sub-thermal switching by effectively amplifying gate control through intrinsic nonequilibrium feedback, offering a more flexible and scalable route to surpass the Boltzmann limit while naturally integrating with in-memory and non-von Neumann architectures~\cite{Silva2022,LopezRichard2022}. However, a comprehensive quantum transport framework integrating these inherent memory mechanisms has remained lacking. In this work, we address this gap by developing a rigorous analytical model based on the Landauer-Büttiker formalism, incorporating memory effects into channel transport to predict and control the subthreshold swing. Our results provide design rules and establish memtransistors as a viable pathway toward overcoming the Boltzmann limit and enabling next-generation ultra-low-power electronics.

Our theoretical model begins with the Landauer–Büttiker formalism, a widely used framework to describe quantum transport in nanoscale conductors~\cite{Datta2017,Sze2021}. For a two-terminal conductor, the net current $I$ flowing through the channel can be expressed as~\cite{Lundstrom2013}
\begin{equation}
I = \frac{2 e}{h} \int \frac{\hbar}{\tau_T(E)} \, \frac{\pi D(E)}{2} \, \bigl(f_2 - f_1\bigr) \, dE ,
\label{eq:LandauerButtiker_2contacts_general}
\end{equation}
where $\tau_T(E)$ is the carrier transit time, dependent on both the transport regime and channel dimensionality, $D(E)$ is the density of states of the channel, and $f_{1}=[1+\exp((E-\mu+e\eta_{ds}V_{ds})/k_BT)]^{-1}$ and $f_{2}=[1+\exp((E-\mu)/k_BT)]^{-1}$ are the Fermi–Dirac distributions of the drain and source reservoirs, respectively, at temperature $T$. The fraction of the total bias that drops along the conductive channel is defined by $\eta_{ds}$. 

\begin{figure}[htbp]
    \centering
\includegraphics[height=0.6\textheight, keepaspectratio]{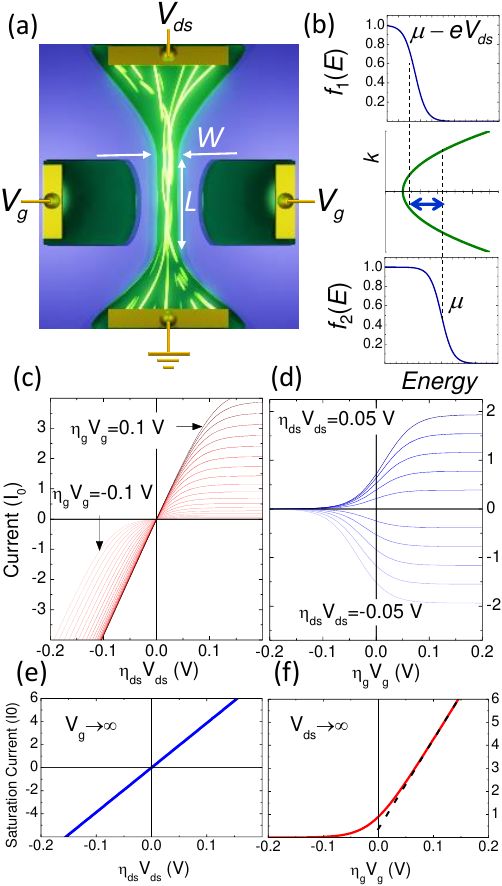}
    \caption{ Baseline characteristics of a 2D diffusive nanotransistor under the near-equilibrium transport approximation.
(a) Schematic of the modeled 2D diffusive field-effect transistor, showing a channel of length $L$ and width $W$, source/drain contacts, and a gate applying voltage $V_g$.
(b) Energy diagram illustrating the principle of thermalization-driven current. An applied bias $V_{ds}$ creates an energy difference between the chemical potentials of the source ($\mu - eV_{ds}$) and drain ($\mu$), opening a Fermi conduction window that enables net current flow.
(c) Calculated output characteristics: current versus bias voltage ($\eta_{ds} V_{ds}$) for various fixed gate voltages ($\eta_g V_g$).
(d) Calculated transfer characteristic: current versus $\eta_g V_g$ for several fixed bias voltages.
(e) Current response in the linear regime   (low $V_{ds}$),  at $V_{g} \rightarrow \infty$.
(f) Saturation current as a function of $\eta_g V_g$ in the high-bias limit ($V_{ds} \rightarrow \infty$), revealing two contrasting approximately linear trends at small $V_g$ (solid red line) and large $V_g$ (dashed line).
All calculations were performed at $T = 300~\mathrm{K}$.
}
    \label{fig:1}
\end{figure}

This description assumes ideal contacts acting as thermal reservoirs, with all inelastic scattering confined to them, while transport in the channel is elastic. Figure~\ref{fig:1}~(a) illustrates the device considered: a 2D nanometric field-effect transistor with channel length $L$ and width $W$, whose charge density is tuned by the gate voltage $V_g$.

Adapting Eq.~(\ref{eq:LandauerButtiker_2contacts_general}) to a two-dimensional diffusive channel under a parabolic band approximation~\cite{Sze2021}, the conduction-band edge shifts as
\begin{equation}
    E_c = E_g - \eta_g e V_g ,
    \label{eq:energy_conductionband}
\end{equation}
with $\eta_g$ the capacitive gate-coupling efficiency. This linear relation reflects a fundamental electrostatic effect: the gate voltage uniformly modulates the potential across the conductive channel, lowering the band bottom for positive gate bias to enhance carrier injection, or raising it for negative bias, thereby depleting carriers and suppressing conduction. The current then takes the form
\begin{equation}
    I = \frac{2 e}{h} \left( \frac{D_n}{\hbar} \frac{W}{L} m^* \right) 
        \int_{E_c}^{\infty} \left[ f_2(E) - f_1(E) \right] \, dE ,
    \label{eq:LandauerButtiker_integralform_dimension_2_type_diffusive}
\end{equation}
where $D_n = \mu_n k_B T/e$ is the diffusion coefficient, $\mu_n$ the mobility, and $m^*$ the effective mass of electrons in the parabolic band~\cite{Datta2017,Sze2021}. The integral represents the so-called Fermi window, which defines the energy range where the nonequilibrium between source and drain gives rise to a net carrier flux~\cite{Datta2017}, as illustrated in Fig.~\ref{fig:1}~(b).
 Solving Eq.~(\ref{eq:LandauerButtiker_integralform_dimension_2_type_diffusive}) yields the general result 
\begin{equation}
    I = \left(\frac{ G_0 k_B T}{e}\right)  
        \ln \!\left[\frac{1+e^{\tfrac{\mu-E_c}{k_B T}}}{1+e^{\tfrac{\mu-E_c-e \eta_{ds} V_{ds}}{k_B T}}}\right] ,
    \label{eq:current_general_2D_diffusive_landauer_buttiker}
\end{equation}
obtained without imposing any specific temperature constraints, with
$
    G_0 = \frac{2 e^2}{h}\left( \frac{\mu_n}{\hbar} \frac{W}{L} m^* \frac{k_B T}{e} \right)
$.

The calculated output and transfer characteristics are shown in Figs.~\ref{fig:1}~(c) and (d), in units of $I_0=G_0 k_BT/e$. In the low-bias regime, defined by $e \eta_{ds} V_{ds} / k_{B} T \ll 1$, the general current expression [Eq.~(\ref{eq:current_general_2D_diffusive_landauer_buttiker})] can be expanded to first order, yielding the linear diffusive current
\begin{equation}
I_{\mathrm{lin}} = \frac{G_{0}}{1 + \exp{\left(\tfrac{E_{c} - \mu}{k_{B}T}\right)}} \, \eta_{ds} V_{ds} ,
\label{eq:CurrentLowBias_LinearEquation}
\end{equation}
with the conductance modulated by the gate potential according to Eq.~(\ref{eq:energy_conductionband}). Finally, in the limit of a fully open channel, where $V_g \rightarrow \infty$, the current state converges to its maximum value with 
$I_{\mathrm{lin}}
  \rightarrow G_{0} \eta_{ds} V_{ds},
$
as shown in Fig.~\ref{fig:1}~(e). This linear-response limit can also be directly derived from Eq.~\ref{eq:current_general_2D_diffusive_landauer_buttiker},valid for any finite and arbitrary value of $V_{ds}$.

In the opposite limit, $V_{ds} \rightarrow \infty$, the current saturates and becomes independent of bias
\begin{equation}
I_{sat} =
I_0 
\ln \left\{
 1 +
 \exp\!\left[
   \frac{ (\mu - E_g) + \eta_g e V_g }{k_B T}
 \right]
\right\},
\label{eq:I_DrainSourcesat_lowGate_finalForm}
\end{equation}
as shown in Fig.~\ref{fig:1}~(f). In this case, two distinct yet linear trends with gate voltage emerge: for small gate voltage, $\eta_{g} e V_g / k_B T \ll 1$, the saturation current (solid red line) reduces to
\begin{equation}
I_{sat} = G_{0}
\left\{ 
  \frac{\eta_{g} V_{g}}{1 + e^{\frac{E_{g} - \mu}{k_{B} T} }}
  + 
  \frac{k_{B} T}{e} 
  \ln \left[  1 + e^{ \frac{\mu - E_{g}}{k_{B} T}  }  \right] 
\right\},
\label{eq:SatDrainSource_cond_low_gate}
\end{equation}
while in the high-gate limit, $\eta_g e V_g / k_B T \gg 1$, it simplifies to
\begin{equation} 
    I_{sat} = G_{0} \left[\eta_{g} V_g+\frac{\left(\mu-E_g\right)}{e}\right],
    \label{eq:I_sat_highGate}
\end{equation}
corresponding to the dashed line in Fig.~\ref{fig:1}~(f). 
Notably, the current saturation at large voltages (both drain-source and gate) is fundamentally determined by the thermalization at the contact reservoirs, considered at thermal equilibrium. Although this framework captures the baseline transport characteristics of our 2D transistor, it cannot yet account for history-dependent effects such as the hysteretic $I$--$V$ curves commonly observed in experiments~\cite{Maier2016,Sangwan2018,Miller2021}.

\begin{figure}[htpb]
    \centering
    \includegraphics[width=\columnwidth]{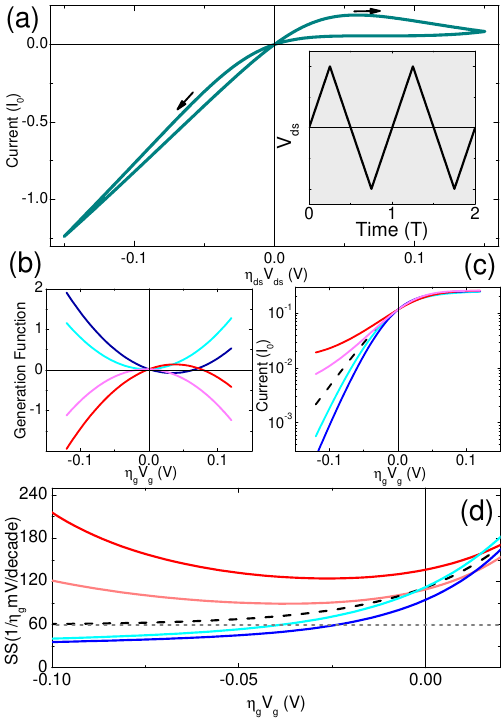}
    \caption{Memory-induced phenomena and sub-thermal switching. 
(a) Calculated $I$--$V$ curve under a triangular voltage sweep (inset), showing the hysteresis loop characteristic of memory effects.  
(b) Gate-voltage-dependent charge generation functions $g(V_g)$ that determine the memory dynamics. 
(c) Corresponding transfer characteristics [lg$(I)$ vs $V_g$], highlighting how memory modifies the subthreshold turn-on compared to the memory-less case (dashed line). 
(d) Subthreshold Swing (SS) as a function of $V_g$, showing that for $\mathrm{d}g/\mathrm{d}V_g < 0$ (blue curves), the model predicts sub-thermal switching with SS well below the 60 mV/decade Boltzmann limit (gray dashed line).
}
    \label{fig:2}
\end{figure}

To account for history-dependent behavior, we introduce a non-equilibrium variation in the trapped charge density, $\delta n$, at the gate interface or within a floating gate~\cite{Maier2016,Miller2021}. This dynamic charge component alters the channel electrostatics by renormalizing the conduction band edge in Eq.~(\ref{eq:energy_conductionband}).
\begin{equation}
    E_{c} =  E_g  - e \, \eta_g V_g + \beta \, \delta n(V_g,V_{ds}),
    \label{eq:energy_conductionband_memory}
\end{equation}
where $\beta$ is a capacitive coupling factor that quantifies the interaction between trapped charge and the channel potential~\cite{Hartmann2017}. For near-equilibrium conditions, the evolution of $\delta n$ follows a relaxation-time approximation
\begin{equation}
  \dfrac{\mathrm{d} \delta n}{\mathrm{d} t} = - \dfrac{\delta n}{\tau_{m}} + g\left(V_{g}, V_{ds}\right),
  \label{eq:memory_rate_eq}
\end{equation}
where $\tau_{m}$ is the characteristic memory relaxation time, and $g(V_g,V_{ds})$ is a generation function describing the trapping and release of carriers~\cite{Silva2022}.  
Following Ref.~\citenum{LopezRichard2022}, the generation function due to thermal activation (or trapping) of carriers can be generalized to include both gate and drain control as
\begin{eqnarray}
    g(V_{g}, V_{ds}) &=&
\frac{F_{0}}{\eta} \Bigg[
-2
+ e^{ -\dfrac{\alpha_{ds}}{1+\alpha_{ds}} \, \dfrac{\gamma_{ds} V_{ds}}{k_{B}T} }
  e^{ -\dfrac{\alpha_{g}}{1+\alpha_{g}} \, \dfrac{\gamma_g V_{g}}{k_{B}T} }  \nonumber
\\
&+& e^{ \dfrac{1}{1+\alpha_{ds}} \, \dfrac{\gamma_{ds} V_{ds}}{k_{B}T} }
  e^{  \dfrac{1}{1+\alpha_{g}} \, \dfrac{\gamma_{g} V_{g}}{k_{B}T} }
\Bigg],
\label{eq:GenFunctionExtended}
\end{eqnarray}
where $F_0$ is the flux weight, $\eta$ an efficiency factor, $\alpha_{ds}$ and $\alpha_g$ characterize the symmetry of trap distributions, and $\gamma_i$ represent the voltage efficiencies for drain and gate control. The interplay of these parameters in Eq.~\eqref{eq:memory_rate_eq} naturally gives rise to pinched hysteretic $I$–$V$ loops, as shown in Fig.~\ref{fig:2}~(a), when the device is subjected to triangular voltage sweeps (inset): an unambiguouis memristive response~\cite{LopezRichard2024a,LopezRichard2024b}.  

Assuming a clear separation of timescales, where the memory relaxation time $\tau_{m}$ is much longer than the carrier transit time, the quasi-static limit yields
\begin{equation}
\delta n = \tau_{m} \, g(V_g, V_{ds}),
\label{eq:memory_deltan}
\end{equation}
linking the non-equilibrium carrier population to a generic generation function $g(V_g, V_{ds})$. This universal formulation captures a wide class of physical processes, such as trapping, detrapping, or defect-assisted recombination, without relying on the complex, material-specific mechanisms often required in ferroelectric models. Using Eq.~(\ref{eq:CurrentLowBias_LinearEquation}), the SS is defined as~\cite{Sze2021}
\begin{equation}
SS = \lim_{\frac{\mu-E_c}{k_BT} \rightarrow -\infty}\left( \dfrac{d \log_{10} I_{lin}}{d V_g} \right)^{-1}.
\label{eq:ss_definition}
\end{equation}
Substituting the renormalized conduction band edge from Eq.~\eqref{eq:energy_conductionband_memory} yields
\begin{equation}
SS = \frac{k_B T \ln(10)}{e} \, \frac{1}{\left(\eta_g - \beta \tau_{m} \dfrac{d g}{d V_g} \right)},
\label{eq:ss_memory}
\end{equation}
the central theoretical result of this work. It demonstrates that nonequilibrium memory dynamics, through the derivative $(dg/dV_g)$, provide an additional and tunable degree of gate control, independent of any material-specific ferroelectric response. In particular, when
\begin{equation}
\frac{dg}{dV_g} < 0,
\label{eq:dg_beatingBoltzmanncondition}
\end{equation}
near the threshold voltage, the device achieves sub-thermal switching, effectively reducing the subthreshold swing below the $60~\mathrm{mV/dec}$ Boltzmann limit by dynamically enhancing gate efficiency. This universal mechanism, illustrated in Figs.~\ref{fig:2}~(b), (c) and (d), highlights how properly engineered nonequilibrium carrier dynamics can replicate and even outperform ferroelectric-like steep-slope behavior in a much simpler and more robust framework.

Note that unlike previous proposals~\cite{Hoffmann2019,Lukyanchuk2022,Chen2024}, the reactive nature of the response: capacitive-like or inductive-like (apparent negative capacitances~\cite{LopezRichard2024a}) is not relevant at all for this recipe to function.
The derivative, $dg/dV_g$ [see Fig.~\ref{fig:2} (b)], represents the sensitivity of carrier generation or trapping rate to changes in the applied gate voltage, and its interpretation depends on the signs of both $g$ and $dg/dV_g$. When $g>0$, the system is in a carrier generation regime: $dg/dV_g>0$ indicates increasing generation with voltage (e.g., due to field-assisted mechanisms), while 
$dg/dV_g<0$ reflects the decay of the generation rate with gate voltage, such as in thermally activated or exponential processes where $g \sim e^{-eV_g/k_BT}$, or may arise from saturation or recombination effects limiting carrier availability. Conversely, when $g<0$, the system is in a trapping regime: 
$dg/dV_g>0$ implies weakening trapping as voltage reduces trap occupancy or activates generation, whereas 
$dg/dV_g<0$ indicates intensifying trapping, likely due to activation of additional traps or stronger immobilization. Thus, $g$ and 
$dg/dV_g$ together reveal the dynamic balance and transitions between generation and trapping in response to voltage changes.

One should note that the explicit dependence of the generation function on $V_{ds}$ and $V_g$ in Eqs.~(\ref{eq:memory_rate_eq}) and (\ref{eq:GenFunctionExtended}) inherently introduces a drain-induced modulation of the current saturation described by Eqs.~(\ref{eq:I_DrainSourcesat_lowGate_finalForm}), (\ref{eq:SatDrainSource_cond_low_gate}), and (\ref{eq:I_sat_highGate}) and an extra gate voltage tuning. This effect represents a potential trade-off between the reduction of the subthreshold swing and the overall transistor performance. Nevertheless, as shown in Fig.~\ref{fig:2}~(c) and (d), this influence remains negligible over a broad range of realistic parameters~\cite{Miller2021}.

In summary, incorporating a phenomenological memory mechanism transforms our baseline transport model into a unified and universal framework capable of describing both the hysteretic $I$--$V$ characteristics commonly observed in nanoscale memtransistors and the emergence of sub-thermal switching. By explicitly coupling nonequilibrium charge-trapping dynamics to quantum transport, the model reveals a general physical principle, independent of material-specific ferroelectric or phase-change effects, through which the Boltzmann limit can be surpassed. This universality makes the approach broadly applicable across device platforms, offering a simpler and more controllable route to steep-slope behavior. Beyond reproducing key experimental signatures, the framework provides clear design guidelines for harnessing intrinsic memory effects as a built-in functionality, enabling energy-efficient, reconfigurable, and multifunctional transistor architectures for next-generation low-power electronics.

\textbf{Acknowledgments} This study was financed in part by the Coordenação de Aperfeiçoamento de Pessoal de Nível Superior - Brazil (CAPES) and the Conselho Nacional de Desenvolvimento Científico e Tecnológico - Brazil (CNPq) Proj. 311536/2022-0 and FAPESP Projs. 2024/09298-7, 2023/05436-3, 2025/00677-8, and 2025/04805-0.

\bibliographystyle{apsrev4-2} 

\bibliography{arXiv.bbl} 

\end{document}